\documentclass[aps,prb,twocolumn,showpacs,superscriptaddress,email]{revtex4}

\usepackage[latin1]{inputenc}
\usepackage{amsmath}
\usepackage[pdftex]{graphicx}
\usepackage{color}

\newcommand{\ket}[1]{\left| #1\right\rangle}

\begin{document}

\title{Dissipation-driven generation of two-qubit entanglement\\
mediated by plasmonic waveguides}

\author{Diego Martín-Cano}
\affiliation{Departamento de Física Teórica de la Materia
Condensada, Universidad Autónoma de Madrid, E-28049 Madrid, Spain}

\author{Alejandro González-Tudela}
\affiliation{Departamento de Física Teórica de la Materia
Condensada, Universidad Autónoma de Madrid, E-28049 Madrid, Spain}

\author{L. Martín-Moreno}
\affiliation{Instituto de Ciencia de Materiales de Aragón (ICMA)\\
and Departamento de Física de la Materia Condensada,\\
CSIC-Universidad de Zaragoza, E-50009 Zaragoza, Spain}

\author{F. J. García-Vidal}
\affiliation{Departamento de Física Teórica de la Materia
Condensada, Universidad Autónoma de Madrid, E-28049 Madrid, Spain}

\author{Carlos Tejedor}
\email[Electronic address: ]{carlos.tejedor@uam.es}
\affiliation{Departamento de Física Teórica de la Materia
Condensada, Universidad Autónoma de Madrid, E-28049 Madrid, Spain}

\author{Esteban Moreno}
\email[Electronic address: ]{esteban.moreno@uam.es}
\affiliation{Departamento de Física Teórica de la Materia
Condensada, Universidad Autónoma de Madrid, E-28049 Madrid, Spain}

\begin{abstract}
We study the generation of entanglement between two distant qubits mediated by the surface plasmons of a metallic waveguide. We show that a V-shaped channel milled in a flat metallic surface is much more efficient for this purpose than a metallic cylinder. The role of the misalignments of the dipole moments of the qubits, an aspect of great importance for experimental implementations, is also studied. A careful analysis of the quantum-dynamics of the system by means of a master equation shows that two-qubit entanglement generation is essentially due to the dissipative part of the effective qubit-qubit coupling provided by the surface plasmons. The influence of a coherent external pumping, needed to achieve a steady state entanglement, is discussed. Finally, we pay attention to the question of how to get information experimentally on the degree of entanglement achieved in the system.
\end{abstract}


\date{\today}

\maketitle

\section{Introduction}
In the last years an intense effort has been made to control and tailor the coupling between quantum emitters and the electromagnetic (EM) field. One major force driving the interest in this research area lies in Quantum Information science, which often requires the transfer of quantum states between matter and light degrees of freedom\cite{nielsen00}. Applications such as quantum teleportation, quantum cryptography, and other two-qubit operations are additionally based on the availability of entangled two-qubit systems. There have been many works analyzing the coupling of qubits provided by the interchange of fermions or bosons\cite{mahan00,lehmberg70} and, in particular, addressing the generation of entanglement due to the coupling to a common bath\cite{braun02,kim02,solenov07,mazzola09}. Within this context, the EM field may constitute the agent needed to prepare a system in a targeted entangled state or to couple two preexisting entangled systems. In a number of proposed schemes qubit-qubit interactions are provided either by coupling to a common EM cavity mode\cite{imamoglu99,plenio99,sorensen03,majer07,laucht10} or, when large separations between the qubits are desired, to a guided mode\cite{kien05,yao09,chenalcubo10}. With independence of the chosen arrangement, the dominant paradigm in quantum-state engineering relies almost exclusively on exploiting the coherent dynamics in order to implement the operations needed for quantum information processing\cite{makhlin01,dutt07}. The traditional view holds that dissipation, being responsible for decoherence, plays only a negative role. However, it has been recently realized that the dissipative dynamics associated with the coupling of the system to external reservoirs can be engineered in such a way that it can drive the system to a desired state encoding the output of a quantum computation\cite{diehl08,verstraete09}. Implementation of such ideas has shown their tremendous potential demonstrating, among other results, the generation of entangled states both in theory\cite{alharbi10,kastoryano11} and experimentally\cite{barreiro11,krauter}.

\begin{figure}[htbp]
\begin{center}
\includegraphics[width=0.99\linewidth,angle=0]{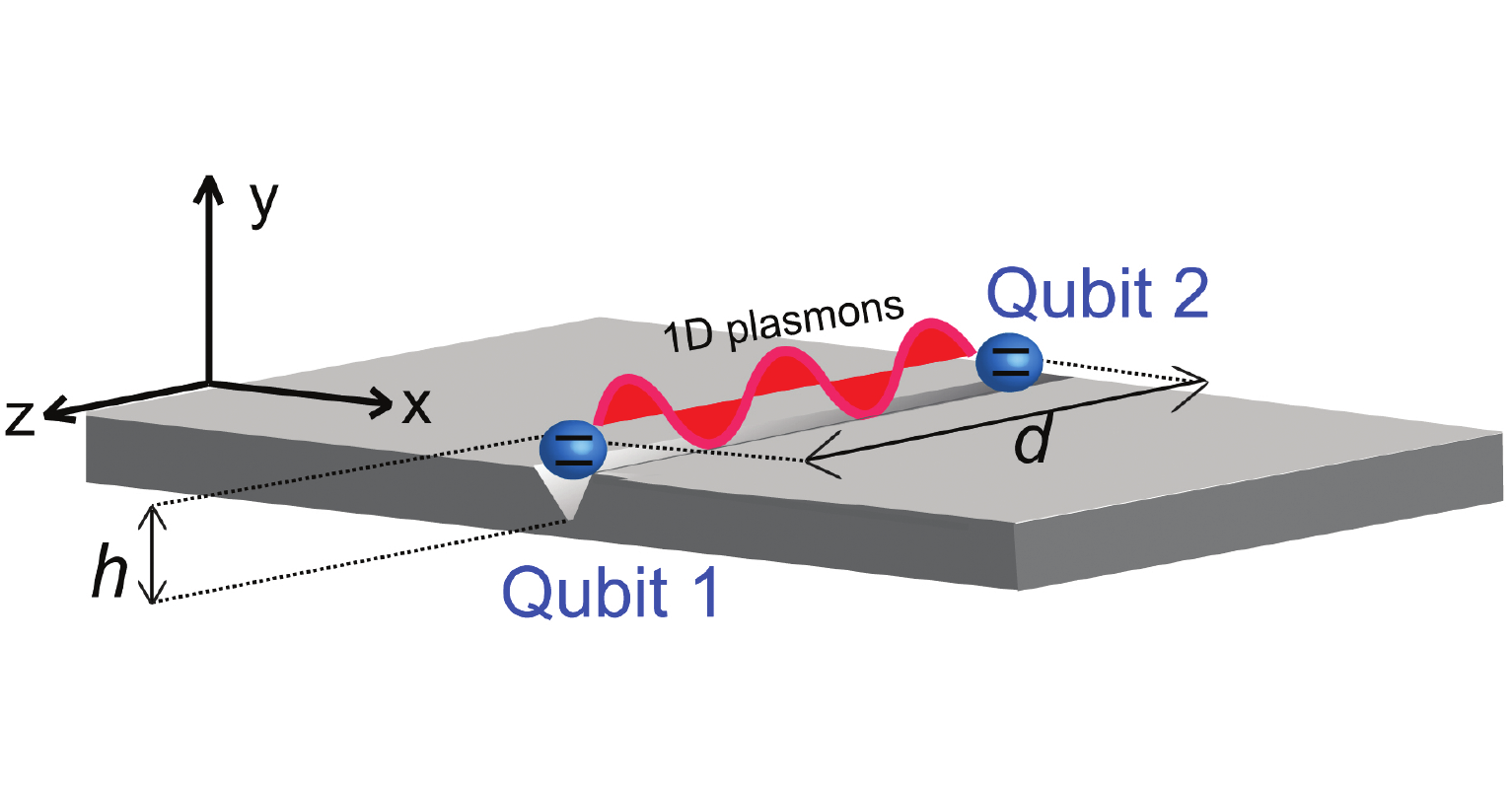}
\end{center}
\caption{ (Color online) Two qubits separated a horizontal distance $d$ are positioned at a vertical distance $h$ from the bottom of a channel waveguide milled in a metallic surface. The plasmon modes supported by the structure mediate the electromagnetic interaction between the qubits.}
\label{system-setup}
\end{figure}

Many structures have been proposed to increase light-matter interaction, including photonic crystal cavities\cite{hughes05,gallardo10} and waveguides\cite{lundhansen08}, photonic nanowires\cite{bleuse11}, and dielectric slot waveguides\cite{quan09}. A crucial requirement for such devices is the enhancement of the EM field, leading to a large Purcell factor, defined as the decay rate of the emitter in the presence of the structure normalized to the decay rate in vacuum. Electric field intensification is favored by a tighter confinement of the EM modes. In connection with this, metallic structures are known to support surface plasmon modes propagating at the metal interface and displaying strong field concentration\cite{raether88}. This modal confinement can reach even the subwavelength level\cite{barnes03}, a feature extensively exploited in plasmonics, \emph{e.g.}, for dense waveguide integration\cite{bozhevolnyi06}. The interaction with surface plasmons has been also employed to control certain properties of quantum emitters, including the decay rate\cite{novotny06}, angular directionality\cite{curto10}, and energy transfer\cite{andrew04,martincano10}. Single plasmon generation\cite{chang06,akimov07} and detection\cite{falk09,heeres10} have been experimentally demonstrated, and the achievements on plasmon transport switching\cite{chang07} and plasmon-assisted qubit-qubit interaction\cite{dzsotjan10}, suggest the on-chip implementation of quantum operations involving qubits in a plasmonic waveguide network. Along this line, we have recently explored the generation of entanglement between two qubits linked by a plasmonic waveguide (PW) consisting of a V-shaped channel milled in a flat metallic surface and operating in the optical regime\cite{gonzaleztudela11a,gonzaleztudela11b} (Fig.~\ref{system-setup}). In our previous work, we showed that the mentioned configuration enables the spontaneous formation of a high degree of entanglement, even for qubit-qubit separations larger than the operating wavelength. In the present paper a more detailed analysis of the two-qubit entanglement generation mediated by plasmons is provided, emphasizing its essential relationship with the dissipative character of the effective two-qubit dynamics. In addition, a more systematic exposition of several aspects of this problem is presented. First, we consider two different waveguide geometries, cylindrical and channel-shaped, analyzing the impact of the waveguide type on the attainable entanglement degree. The role of dipole moment misalignments is also assessed, which is of great importance for experimental implementations due to the difficulty in the controlled emitter positioning. The influence of the intensity of the coherent external pumping, needed to achieve a steady state entanglement, is discussed. We also pay attention to the question of how to detect experimentally the degree of entanglement achieved in the system by measuring cross terms of a second order coherence function. Finally, we study the effect of pure dephasing produced by non-radiative mechanisms.

\begin{figure}[htbp]
\begin{center}
\includegraphics[width=0.99\linewidth,angle=0]{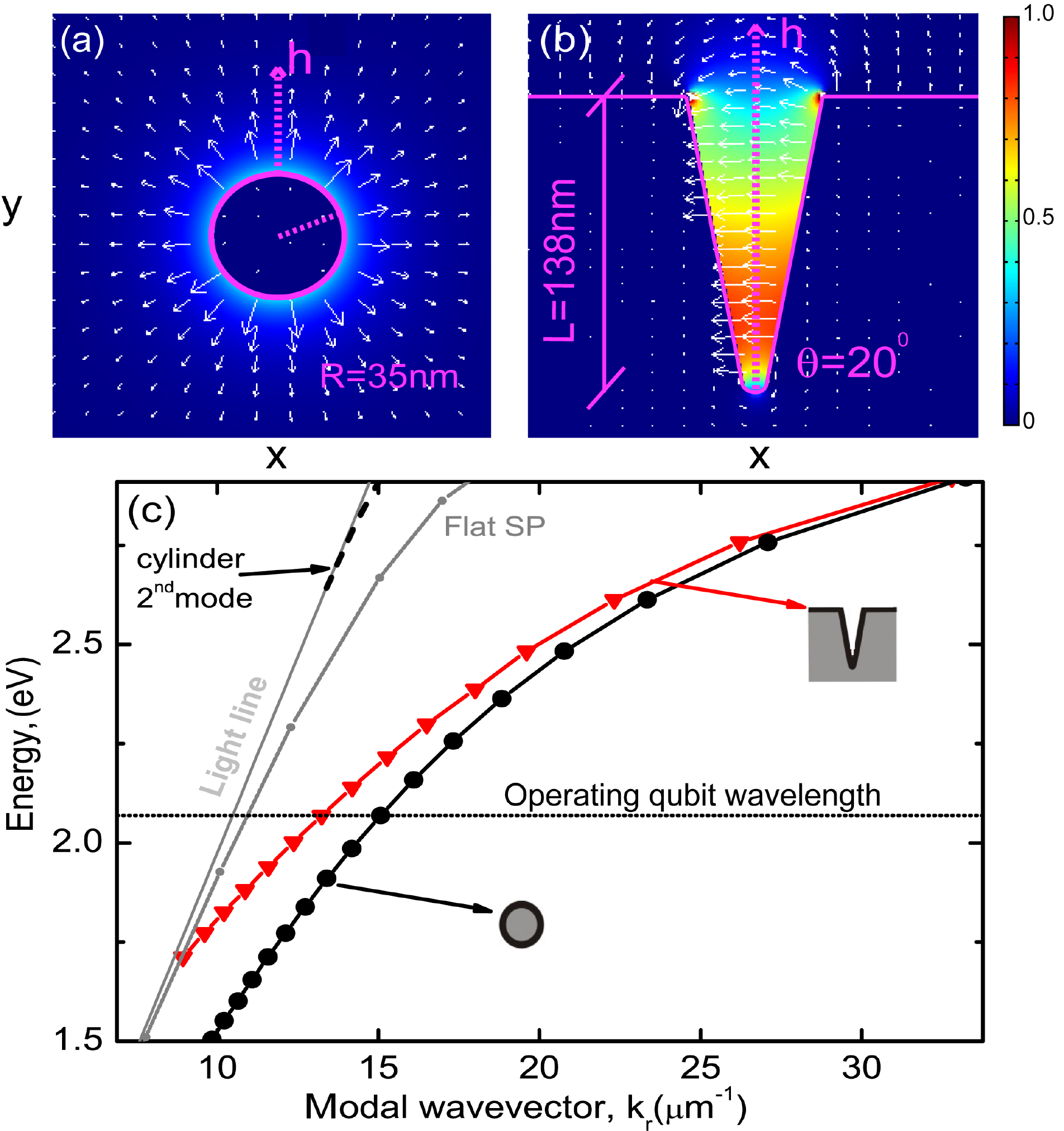}
\end{center}
\caption{ (Color online) Transverse cross section of a cylindrical nanowire (a) and a channel waveguide (b). The color scale in  (a) and (b) renders the transverse electric field amplitude of the supported plasmonic modes, and the arrows show the electric field polarization. (c) Dispersion relation for the fundamental mode of the cylindrical nanowire (black circles) and channel waveguide (red triangles). The vacuum light line, the dispersion relation of a plasmon on a flat silver surface, and that of the second mode supported by the cylinder are also plotted.}
\label{fields-dispersion}
\end{figure}

The rest of the paper is organized in five sections: Sec.~II contains a short description of the setup and the main PWs characteristics. In Sec.~III we summarize briefly the formalism of the master equation governing the effective two-qubit dynamics. In addition, the measures of entanglement and correlation are recalled. Section~IV describes how the classical Green's tensor and the associated coupling constants entering the master equation are computed. Then, the influence of various aspects such as waveguide type, emitter position, and dipole moment orientation are analyzed. Once these results are available, the generation of entanglement with or without external laser pumping is discussed in Sec.~V, and its relation with the dissipative dynamics is highlighted. We also study the relationship between entanglement and photon-photon correlations and the influence of the pumping rate and pure dephasing. Section~VI is devoted to the conclusions.

\section{System description and plasmonic waveguides characteristics}
The system analyzed in this paper consists of two identical quantum emitters positioned in closed proximity to a metallic waveguide (Fig.~\ref{system-setup}), in such a way that their EM interaction is dominated by the plasmonic modes supported by the quasi one-dimensional structure. The emitters, which could be atoms, molecules, quantum dots, or nitrogen-vacancy centers in diamond, will be modeled as two-level systems, with a transition frequency $\omega_0$ corresponding to an emission wavelength $\lambda=600\,\textmd{nm}$. A point-emitter approach is assumed because it contains all the main physics of the problem without involving a detailed description of each qubit, which can be cumbersome for large molecules or quantum dots\cite{andersen11}. In order to determine the influence of the PW geometry, we consider two different metallic structures: the first is a cylindrical nanowire and the second a channel waveguide (the case depicted in Fig.~\ref{system-setup}). These waveguide types have been previously fabricated and successfully demonstrated for dense waveguiding\cite{bozhevolnyi06} and single plasmon generation\cite{akimov07}. The exact geometry of both structures is detailed in panels (a) and (b) of Fig.~\ref{fields-dispersion}. The radius of the cylindrical nanowire is $R=35\,\textmd{nm}$, the depth of the V-shaped groove is $L=138\,\textmd{nm}$, and its angle is $\theta=20^{\circ}$. The considered metal is silver, whose electric permittivity at the mentioned wavelength is\cite{rodrigo08} $\varepsilon=\varepsilon_{\textmd{r}}+i\varepsilon_{\textmd{i}}=-13+i0.8$. The geometric parameters of both structures have been chosen so that, at the operating wavelength, only one mode is relevant and the propagation length is identical for cylinders and channels. The channel waveguide is single-moded and the cylinder supports two modes but the second one (black dashed line), being extremely close to the light line, is very much extended in the transverse cross plane and will not play a relevant role in what follows. Since the qubit-qubit interaction will be mediated by the plasmonic modes, having identical propagation length ensures a meaningful comparison of the results obtained with both PWs. The propagation length is $\ell=[2 k_{\textmd{i}}]^{-1}=1.7\,\mu\textrm{m}$, $k_{\textmd{i}}$ being the imaginary part of the (complex) modal wave vector, $k=k_{\textmd{r}}+ik_{\textmd{i}}$. The dispersion relation for both PWs is rendered in Fig.~\ref{fields-dispersion}(c) and it is observed that the curve corresponding to the cylinder (black circles) lies to the right of that corresponding to the channel (red triangles), implying that the EM field of the former is more tightly confined. This is confirmed by a comparison of panels (a) and (b), where the transverse electric field modal profiles and polarizations are plotted. For both waveguides the modal size is deep-subwavelength. In spite of the fact that the electric field of both structures includes transverse and longitudinal components, the former dominate by a factor of about 10. For this reason it will be later advantageous to orient the emitters parallel to the transverse plane.

\section{Two-qubit dynamics, entanglement, and correlation}
In this section the tools required to determine the quantum state of two qubits and their entanglement degree are reviewed. The evolution of the two qubits in interaction with the EM field supported by a plasmonic waveguide can be represented using a Green's tensor approach to macroscopic quantum electrodynamics\cite{knoll01,dung02,dzsotjan10}. One important advantage of this method is that all magnitudes describing the coupling between the qubits and the EM field can be obtained from the classical Green's tensor appropriate for the corresponding structure. Within this approach, the Hamiltonian for the system in the presence of a dispersive and absorbing material is written in the electric dipole approximation as
\begin{equation}
\label{detailedhamiltonian}
\begin{aligned}
\hat{H}=\int d^{3}\mathbf{r}\int^{\infty}_{0}d\omega\,\hbar\omega\, \mathbf{\hat{f}}^{\dag}(\mathbf{r},\omega)\mathbf{\hat{f}(r,\omega)}
&+\sum_{i=1,2}\hbar\omega_{0}\,\hat{\sigma}^{\dag}_{i}\hat{\sigma}_{i}\\
-\sum_{i=1,2}\int^{\infty}_{0} d\omega [\mathbf{\hat{d}}_{i}\mathbf{\hat{E}}(\mathbf{r}_{i},\omega)+\textmd{h.c.}].
\end{aligned}
\end{equation}
Here $\mathbf{\hat{f}}^{\dag}$ and $\mathbf{\hat{f}}$ represent the bosonic fields in the medium with absorption, which play the role of the fundamental variables of the electromagnetic field and the dielectric medium. $\hat{\sigma}_{i}^{\dag}$ is the $i$-qubit rising operator, $\mathbf{r}_i$ its spatial position, $\omega_0$ is the transition frequency,  and $^{\dag}$ stands for the adjoint operation. The interaction term includes the dipole moment operator $\mathbf{\hat{d}}_{i}=\mathbf{d}_{i}\hat{\sigma}_{i}+\mathbf{d}^{*}_{i}\hat{\sigma}^{\dag}_{i}$,
where $\mathbf{d}_{i}$ is the dipolar transition matrix element and $^{*}$ denotes complex conjugation. In addition,
\begin{equation}
\label{Efieldoperator}
\mathbf{\hat{E}}(\mathbf{r},\omega)=i\sqrt{\frac{\hbar}{\pi\epsilon_{0}}}\frac{\omega^{2}}{c^{2}}\int d^{3}\mathbf{r'}\sqrt{\varepsilon_{\textmd{i}}(\mathbf{r'},\omega)}\mathbf{G(r,r',\omega)}\mathbf{\hat{f}}(\mathbf{r'},\omega)
\end{equation}
is the electric field operator. Notice the explicit appearance of the Green's tensor $\mathbf{G(r,r',\omega)}$, which satisfies the classical Maxwell equations for an infinitesimal dipole source located at the spatial position $\mathbf{r'}$. Physically, the Green's tensor carries the electromagnetic interaction from the spatial point $\mathbf{r'}$ to $\mathbf{r}$.

This hamiltonian description is very powerful but, as a matter of fact, it contains too much detail for the purpose of this paper. The following simpler description, that derives from the previous one, will be employed here. To determine the entanglement of the two qubits induced by their EM interaction, we only need an equation governing the dynamics of the reduced density matrix $\mathbf{\hat{\rho}}$ corresponding to the two-qubit system. Such a representation of the dynamics is obtained from Eqs.~(\ref{detailedhamiltonian}) and (\ref{Efieldoperator}) by tracing out the EM degrees of freedom. The corresponding master equation, whose derivation can be found in Refs.~\onlinecite{ficek02,dung02}, reads as follows:
\begin{equation}
\label{masterequation}
\frac{\partial \hat{\rho}}{\partial t}=- \frac{i}{\hbar}[\hat{H}_{\textmd{s}},\hat{\rho}]-\frac{1}{2}\sum_{i,j}\gamma_{ij}(\hat{\rho}\mathbf{\hat{\sigma}}^{\dag}_{i}\mathbf{\hat{\sigma}}_{j}+\mathbf{\hat{\sigma}}^{\dag}_{i}\mathbf{\hat{\sigma}}_{j}\hat{\rho}-2\mathbf{\hat{\sigma}}_{j}\hat{\rho}\mathbf{\hat{\sigma}}^{\dag}_{i}),
\end{equation}
where the hamiltonian included in the coherent part of the dynamics is
\begin{equation}
\label{hamiltonianS}
\hat{H}_{\textmd{s}}=\sum_{i}\hbar (\omega_0+\delta_i)\,\hat{\sigma}^{\dag}_{i}\hat{\sigma}_{i}+\sum_{i\neq j}\hbar g_{ij}\,\hat{\sigma}^{\dag}_{i}\hat{\sigma}_{j}.
\end{equation}
The interpretation of the various constants appearing in Eqs.~(\ref{masterequation}) and (\ref{hamiltonianS}) is the following. The Lamb shift, $\delta_i$, is due to the qubit EM self-interaction in the presence of the PW. At optical frequencies, for qubit-metal distances larger than about $10\,\textmd{nm}$, $\delta_i$ is very small\cite{novotny06book,hohenester08} and will be neglected in what follows. The level shift induced by the dipole-dipole coupling is given by $g_{ij}$, and can be evaluated approximately from
\begin{equation}
\label{gij}
g_{ij}= \frac{\omega_{0}^{2}}{\hbar\varepsilon_{0}c^{2}}\mathbf{d}_{i}^{*}\,\textmd{Re}\mathbf{G}(\mathbf{r}_{i},\mathbf{r}_{j},\omega_{0})\,\mathbf{d}_{j}.
\end{equation}
Finally, the parameters in the dissipative (noncoherent) term of Eq.~(\ref{masterequation}) are given approximately by
\begin{equation}
\label{gammaij}
\gamma_{ij}=
\frac{2\omega_{0}^{2}}{\hbar\varepsilon_{0}c^{2}}\mathbf{d}_{i}^{*}\,\textmd{Im}\mathbf{G}(\mathbf{r}_{i},\mathbf{r}_{j},\omega_{0})\,\mathbf{d}_{j},
\end{equation}
and represent the decay rates induced by the self ($\gamma_{ii}$) and mutual ($\gamma_{ij}$) interactions. Expressions (\ref{gij}) and (\ref{gammaij}) are obtained by integration of the EM field Green's function in the frequency domain\cite{dung02}. To reach the result that the coherent and incoherent contributions to the coupling are proportional to the real and imaginary parts of the Green's function, respectively, the Kramers-Kronig relation between the real and imaginary parts of the Green's function is used\cite{dung02,dzsotjan}. In deriving the master equation a Born-Markov approximation is applied, valid for weak qubit-EM field interaction and broadband PWs. Let us remark that, as mentioned above, both $g_{ij}$ and $\gamma_{ij}$ can be extracted from the knowledge of $\mathbf{d}_{i}$ and the classical Green's tensor in the presence of the PW. The dipole moment can be inferred from the measurement of the decay rate of one qubit in vacuum, whose Green's tensor is well known. Up to this point it has been assumed that both dipoles have equal frequencies, but we would like to remark that the formalism is a good approximation when the frequencies are unequal but sufficiently close to each other. In this regard various criteria can be mentioned. According to Refs.~\onlinecite{ficek02} and~\onlinecite{milonni75} the frequency difference should be much smaller than the average frequency, whereas Dung and coworkers\cite{dung02} state that the frequency difference should be smaller than the typical frequency scale for which the Green´s tensor displays a significant variation. We have checked that both criteria are fulfilled for dipoles whose emission wavelengths are in the range of $600\,\textmd{nm}$ and differ by less than about ten nanometers.

\begin{figure}[htbp]
\begin{center}
\includegraphics[width=0.99\linewidth,angle=0]{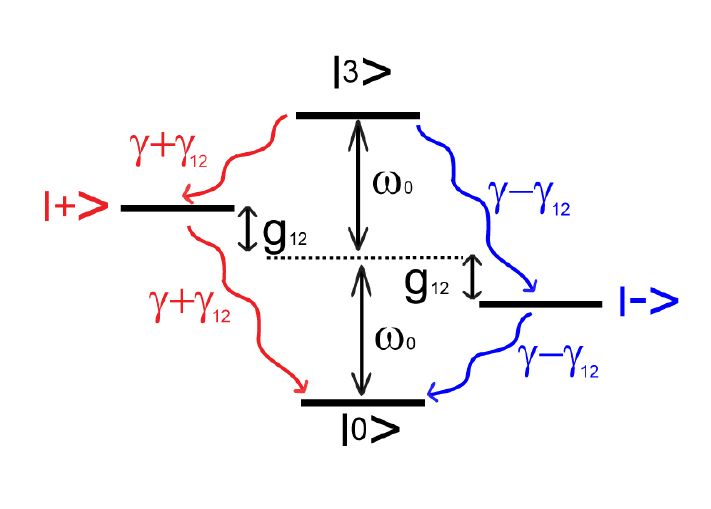}
\end{center}
\caption{ (Color online) Scheme of levels for two identical qubits located at equivalent positions with respect to the PW and with identical orientations ($\gamma_{11}=\gamma_{22}=\gamma$ and $\gamma_{12}=\gamma_{21}$).}
\label{levelscheme}
\end{figure}

To solve Eq.~(\ref{masterequation}) a basis for the vector space corresponding to the two-qubit system has to be chosen. A convenient basis that makes $\hat{H}_{\textmd{s}}$ diagonal is formed by the following states: $\ket{3}=\ket{e_1 e_2}$, $\ket{0}=\ket{g_1 g_2}$, and $\ket{\pm}=\frac{1}{\sqrt{2}}(\ket{g_1 e_2}\pm\ket{e_1 g_2})$, where $\ket{g_i}$ ($\ket{e_i}$) labels the ground (excited) state of the $i$-qubit. Using this basis the evolution of the diagonal elements of Eq.~(\ref{masterequation}) is given by
\begin{eqnarray}
\label{masterequationbasis}
\nonumber\dot{\rho}_{33}(t)&=&-2\gamma \rho_{33}(t)\\
\nonumber\dot{\rho}_{++}(t)&=&(\gamma+\gamma_{12})\rho_{33}(t)-(\gamma+\gamma_{12})\rho_{++}(t)\\
\nonumber\dot{\rho}_{--}(t)&=&(\gamma-\gamma_{12})\rho_{33}(t)-(\gamma-\gamma_{12})\rho_{--}(t)\\
\nonumber\dot{\rho}_{00}(t)&=&(\gamma+\gamma_{12})\rho_{++}(t)+(\gamma-\gamma_{12})\rho_{--}(t),\\
\end{eqnarray}
where it has been assumed that the positions and orientations of the two qubits in their respective planes transverse to the PW are identical, so that $\gamma_{11}=\gamma_{22}=\gamma$ and $\gamma_{12}=\gamma_{21}$. The diagonal character of $\hat{H}_{\textmd{s}}$ in the above mentioned basis and the interpretation of Eqs.~(\ref{masterequationbasis}) is depicted in Fig.~\ref{levelscheme}, including the level scheme and the collective decay rates induced by the coupling to the EM field. Once these decay rates are evaluated in Sec. IV, the generation of entanglement will be elucidated with the help of this diagram. Notice that the qubit-qubit dissipative coupling induces modified collective decay rates $(\gamma+\gamma_{12})$ and $(\gamma-\gamma_{12})$ which, for particular conditions to be detailed in Sec.~V, give rise to subradiant and superradiant states.

Up to now we have assumed that the system evolves without the influence of any external agent. As a consequence, the upper levels in Fig.~\ref{levelscheme} become eventually depopulated and the ground level $\ket{0}$, an unentangled state, is reached. To prevent this situation, the decays can be compensated by externally pumping the two qubits,
thus maintaining the system in an excited steady state. In cavity quantum electrodynamics, the usual situation is that of incoherent pumping\cite{delValle07,delValle11a} due to the practical difficulties of coherently exciting qubits which are embedded in a cavity. However, our system is geometrically simpler and one can produce a coherent pumping by means of a laser whose frequency, $\omega_{\textmd{L}}$, is close to resonance with the frequency of the qubits\cite{gonzaleztudela11a,gonzaleztudela11b}. The description of this new element requires the inclusion of an additional term in the hamiltonian of Eq.~(\ref{hamiltonianS}):
\begin{equation}
\label{hamiltonianL}
\hat{H}_{\textmd{L}}=-\frac{1}{2}\sum_{i}[\hbar \Omega_i \, \hat{\sigma}^{\dag}_{i}
e^{i\omega_{\textmd{L}}t}+ \textmd{h.c.}].
\end{equation}
Here the strength and phase of the laser are characterized by the Rabi frequencies $\Omega_i=\mathbf{d}_{i}\mathbf{E}_{\textmd{L}}\,e^{i\mathbf{k}_{\textmd{L}}\mathbf{r}_{i}}/\hbar$, where $\mathbf{E}_{\textmd{L}}$ and $\mathbf{k}_{\textmd{L}}$ are the amplitude and wave vector of the driving laser field, respectively. In the most general case, the determination of the density matrix $\mathbf{\hat{\rho}}(t)$ requires the numerical integration of Eq.~(\ref{masterequation}) with appropriate initial conditions\cite{ficek90}. When the system is pumped, the steady state solution can be obtained by setting $\dot{\hat{\rho}}=0$ and solving the corresponding linear equations.

In both scenarios (pumped and non-pumped), once the density matrix $\hat{\rho}(t)$ is known it is possible to compute various magnitudes of interest, such as those quantifying the two-qubit entanglement, or first and second order coherence functions, which are directly related to measurable properties. Regarding the quantification of entanglement, there are several alternatives but all of them are related to each other for a bipartite system\cite{horodecki09}. In this paper we make use of the Concurrence\cite{wootters98}, which ranges from 0 for unentangled states to 1 for maximally entangled states, and is defined as follows: $C\equiv[\mathrm{max}\{0,\sqrt{\lambda_1}-\sqrt{\lambda_2}-\sqrt{\lambda_3}- \sqrt{\lambda_4}\}]$, where $\{\lambda_1, \lambda_2, \lambda_3, \lambda_4 \}$ are the eigenvalues
of the matrix $\rho T \rho^*T$ in decreasing order  (the operator $T$ is  $\sigma_{y}\otimes\sigma_{y}$, $\sigma_{y}$ being the Pauli matrix). Typical measurable magnitudes include two-times coherence functions\cite{walls94,scully_book02}. Their calculation is cumbersome but completely standard, since the quantum regression theorem\cite{walls94,scully_book02} establishes that any two-times coherence function obeys the same dynamics as that of the density matrix $\hat{\rho}(t)$, \emph{i.e.}, Eq.~(\ref{masterequation}). As it will be discussed in Sec.~V, entanglement is related with coherence functions at zero delay. These are measurable by means of a Hanbury Brown-Twiss-like experiment detecting photon-photon correlations in the emission produced by the de-excitation of the qubits. One advantage of zero delay correlations is that their calculation is simple because it does not require the use of the quantum regression theorem.

\section{Computation of the Green's tensor, decay rates, and dipole-dipole shifts}
In this section we compute the Green's tensor corresponding to the PWs described in Sec.~II. This tensor encapsulates the influence of the inhomogeneous environment and is required for the determination of the decay rates, $\gamma_{ij}$, and dipole-dipole shifts, $g_{ij}$, appearing in the master equation.

\subsection{Purcell factor}
For very symmetric structures such as metallic planes\cite{sondergaard04} or cylinders\cite{sondergaard01} analytic expressions for the Green's tensor are available, but for the less symmetric case of a channel PW numerical simulations are necessary. Using the relationship\cite{novotny06book}
\begin{equation}
\label{E-G}
\mathbf{E}(\mathbf{r})=\omega^2\mu_0 \mathbf{G}(\mathbf{r},\mathbf{r'})\mathbf{d},
\end{equation}
the Green's tensor can be inferred if the electric field $\mathbf{E}(\mathbf{r})$ in position $\mathbf{r}$ radiated by a classical oscillating electric dipole $\mathbf{d}$ at the source position $\mathbf{r'}$ is known. We compute the EM field excited by the dipole source with the finite element method (FEM)\cite{jin02,chen10} using commercial software (COMSOL). The point dipole is modeled as a linear harmonic current of length $l$, intensity $I_0$, and orientation given by the unit vector $\mathbf{n}$. The associated dipole moment is\cite{chen10} $\mathbf{d}=(i I_0 l / \omega) \mathbf{n}$ and, to satisfy the dipole approximation, the length $l$ is kept very short in comparison with the emission wavelength ($l=\lambda/330$). To model infinitely long PWs the spatial domain of interest is properly terminated with Perfect Matching Layers, which absorb the outgoing electromagnetic waves with negligible reflection. The size of the simulation domain is of the order of $30\lambda^3$. A non uniform mesh is employed where the typical element sizes are chosen to satisfy the following criteria: $\sim\lambda/300$ in the dipole neighborhood, $\sim\lambda/40$ at the waveguide metal interfaces, $\sim\lambda/12$ at the planar metal interface surrounding the channel, and $\sim\lambda/4$ in vacuum away from the source.

\begin{figure}[htbp]
\begin{center}
\includegraphics[width=0.99\linewidth,angle=0]{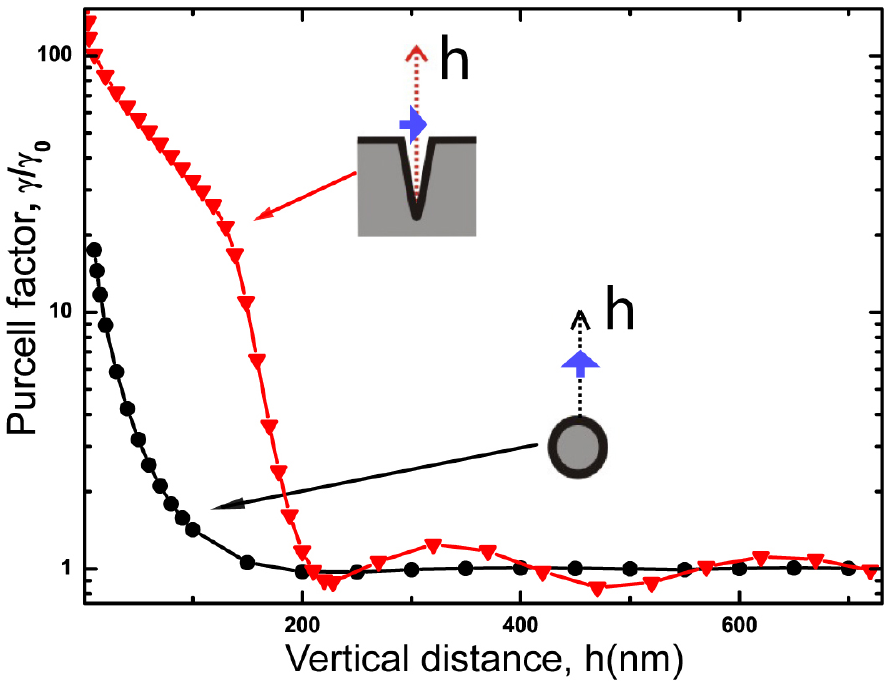}
\end{center}
\caption{ (Color online) Purcell factor ($\gamma/\gamma_0$) versus vertical height $h$ of the emitter along the lines displayed in the insets. Cylinder (black circles) and channel (red triangles).}
\label{purcellfactor}
\end{figure}

Following the explained procedure we now evaluate Eq.~(\ref{gammaij}) to compute the total decay rate, $\gamma=\gamma_{11}$, of one qubit in the presence of a PW. This magnitude appears in Eq.~(\ref{masterequationbasis}) setting the time scale of the dynamics. The Purcell factor, $\gamma/\gamma_0$, is plotted in Fig.~\ref{purcellfactor} as a function of the vertical distance $h$ between the PW and the qubit along the vertical lines displayed in the insets ($\gamma_0$ denotes the decay rate in vacuum). To achieve optimal coupling the dipole is aligned parallel to the field polarization, \emph{i.e.}, vertically for the cylindrical waveguide and horizontally for the channel. The Purcell factor is strongly enhanced when the emitter is very close to the metal surface ($h\rightarrow0$). This effect is more pronounced for the channel, due to a higher electric field when the emitter lies at the bottom of the groove. The curve corresponding to the channel waveguide displays distinct oscillations for large $h$. These are the result of constructive and destructive interference of the direct field and the field reflected mainly at the flat metallic interface surrounding the channel.

\subsection{$\beta$ factor}
\begin{figure}[htbp]
\begin{center}
\includegraphics[width=0.99\linewidth,angle=0]{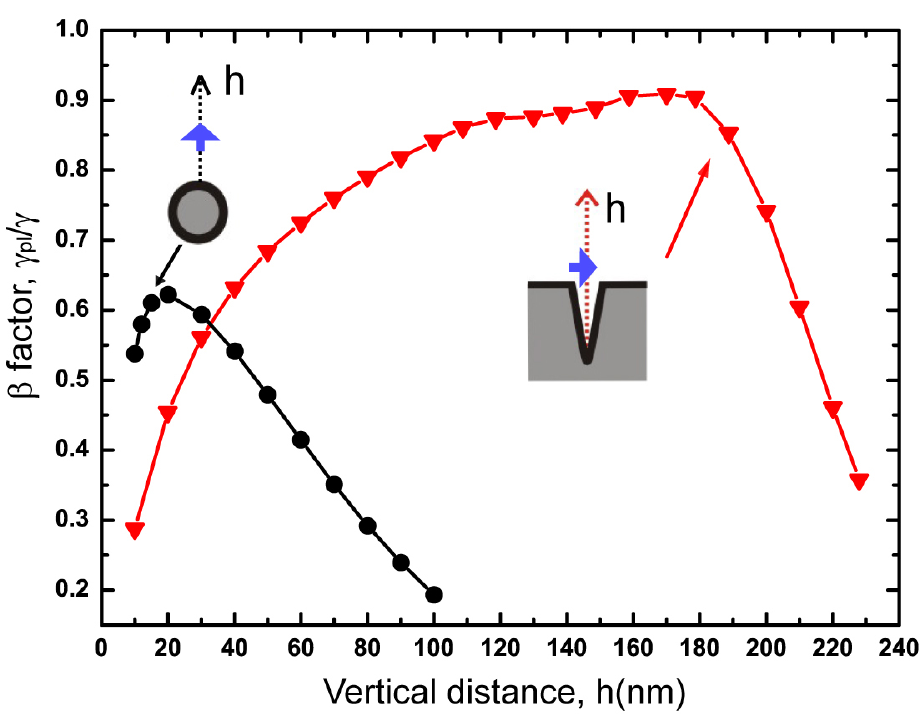}
\end{center}
\caption{ (Color online) Beta factor ($\gamma_{\textrm{pl}}/\gamma$) versus vertical height $h$ of the emitter along the lines displayed in the insets. Cylinder (black circles) and channel (red triangles).}
\label{betafactor}
\end{figure}

The total dipole emission that we have just presented can be either radiated to vacuum, non-radiatively absorbed in the metal, or coupled to guided modes\cite{ford84,chen10}. It is thus customary to express the total decay rate as the sum of those three contributions, $\gamma=\gamma_{\textmd{r}}+\gamma_{\textmd{nr}}+\gamma_{\textmd{pl}}$. The photons absorbed in the metal and most photons radiated to vacuum do presumably not contribute to the qubit-qubit coupling. It will therefore be interesting to compute the decay rate to plasmons, $\gamma_{\textmd{pl}}$, and the fraction of all emission that is coupled to plasmons, $\beta=\gamma_{\textmd{pl}}/\gamma$. As will be shown later, these magnitudes play a dominant role in the qubit-qubit interaction for appropriate qubit-PW vertical distance. In a similar way to the above mentioned total decay rate decomposition, the total Green's tensor can be separated as the sum of several terms corresponding to the three emission channels. In order to compute $\gamma_{\textmd{pl}}$, the plasmon contribution to the Green's tensor is required, which is given by\cite{snyder83,collin90}
\begin{equation}
\label{Gplasmon}
\mathbf{G}_{\textrm{pl}}(\mathbf{r},\mathbf{r}')=\frac{i\,\,\mathbf{E}^{\textrm{t}}(\mathbf{r}^{\textrm{t}})\otimes\mathbf{E}^{\textrm{t}}(\mathbf{r'}^{\textrm{t}})}{2\omega\mu_{0}\int_{S_{\infty}}dS\,\mathbf{u}_z(\mathbf{E}^{\textrm{t}}\times\mathbf{H}^{*\textrm{t}})}\,\,e^{ik(z-z')}.
\end{equation}
The occurrence of the exponential factor $e^{ik(z-z')}$ mirrors the quasi one-dimensional character of the PW-mediated interaction. The lateral extension of the plasmon is taken into acount by $\mathbf{E}^{\textrm{t}}(\mathbf{r}^{\textrm{t}})$ and $\mathbf{H}^{\textrm{t}}(\mathbf{r}^{\textrm{t}})$, which are the transverse EM fields corresponding to the mode supported by the PW [Figs.~\ref{fields-dispersion}(a) and (b) display the transverse electric field] and are evaluated at the transverse position $\mathbf{r}^{\textrm{t}}=(x,y)$. $S_{\infty}$ is the (infinite) transverse area, $\mathbf{u}_z$ is a longitudinal unit vector, and $\otimes$ denotes the tensor product. The derivation of Eq.~(\ref{Gplasmon}) assumes that the mode propagates towards the right ($z>z'$) and its absorption is not too high. To be more precise, Eq.~(\ref{Gplasmon}) is the transverse part of the Green's tensor, which is the relevant part since we will only consider transversely oriented dipole moments. The modal fields entering Eq.~(\ref{Gplasmon}) are obtained by FEM numerical simulation of the corresponding eigenvalue problem\cite{chen10,jung07}. Inserting Eq.~(\ref{Gplasmon}) in the expression for the decay rate (Eq.~\ref{gammaij}) we obtain
\begin{equation}
\label{gamma-pl}
\gamma_{ij,\,\textrm{pl}}=\frac{\omega\,\, [\mathbf{d}_i \mathbf{E}^{\textrm{t}}(\mathbf{r}_{i}^{\textrm{t}})]\,[\mathbf{d}_j \mathbf{E}^{\textrm{t}}(\mathbf{r}_{j}^{\textrm{t}})]}{\hbar \int_{S_{\infty}}dS\,\mathbf{u}_z(\mathbf{E}^{\textrm{t}}\times\mathbf{H}^{*\textrm{t}})} e^{-k_{\textrm{i}}(z-z')}\cos[k_{\textrm{r}}(z-z')],
\end{equation}
which, for $\mathbf{r}_i=\mathbf{r}_j$ and $\mathbf{d}_i=\mathbf{d}_j$, becomes the plasmonic decay rate, $\gamma_{\textrm{pl}}$. This expression clarifies that $\gamma_{\textrm{pl}}$ is largest when the emitter is positioned at the field maximum and aligned with the field polarization. Once $\gamma$ and $\gamma_{\textrm{pl}}$ have been determined, we can plot the $\beta$ factor as a function of the vertical distance $h$ between the PW and the qubit (Fig.~\ref{betafactor}). The general behavior is similar for both the cylindrical and channel PWs. First, the $\beta$ factor is very low for small emitter-PW distance, in sharp contrast to what is observed for the Purcell factor in Fig.~\ref{purcellfactor}. The explanation is that $\gamma_{\textrm{nr}}$ behaves as $h^{-3}$, where $h$ is the qubit-metal distance\cite{ford84}, being the dominant contribution to $\gamma$ for $h\rightarrow0$ and effectively quenching the plasmon emission. For intermediate $h$ the plasmonic decay dominates and $\beta$ attains a maximum. Finally, for large $h$ the emitter is outside the reach of the plasmon mode and the unbounded radiative modes have a larger weight leading to a decrease in $\beta$. Nevertheless, the precise behavior of $\beta$ is not identical for both PWs. Channels display a higher maximum than cylinders (0.91 at $h=160\,\textmd{nm}$ versus 0.62 at $h=20\,\textmd{nm}$, respectively) and, in addition, the maximum is broader for channels than for cylinders ($\beta$ deviates less than a 10\% of the maximum value within a $h$-range of $\Delta h=100\,\textmd{nm}$ for channels and of only $\Delta h=30\,\textmd{nm}$ for cylinders). These features make channels a more attractive structure to enhance the interaction mediated by plasmons, in the range of parameters explored.

\subsection{Dipole-dipole shift and decay rates}
\begin{figure}[htbp]
\begin{center}
\includegraphics[width=0.99\linewidth,angle=0]{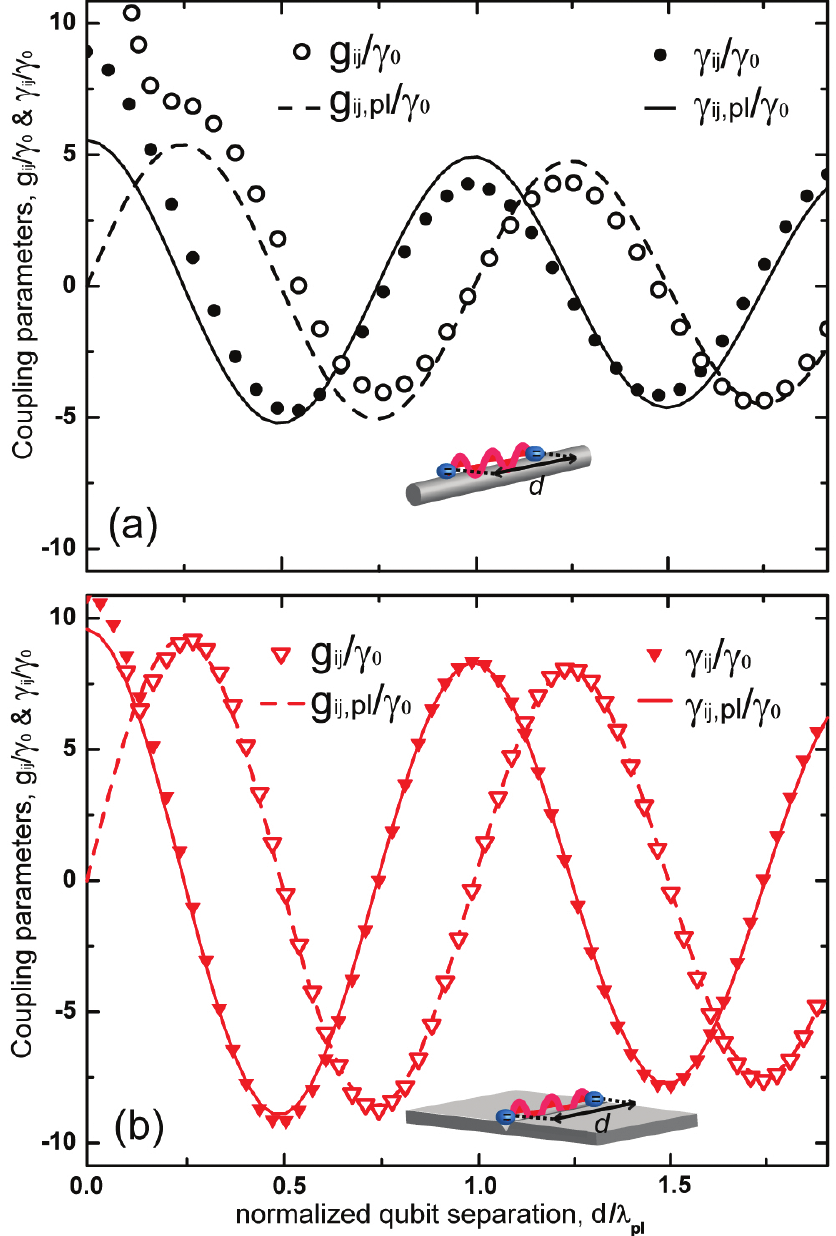}
\end{center}
\caption{ (Color online) Comparison of the exact coupling parameters ($\gamma_{ij}$, $g_{ij}$) with their plasmonic contributions ($\gamma_{ij,\,\textrm{pl}}$, $g_{ij,\,\textrm{pl}}$), as a function of the qubit-qubit horizontal separation normalized to the plasmon modal wavelength, $d/\lambda_{\textrm{pl}}$. All parameters are normalized to the vacuum decay rate $\gamma_0$. (a) Cylindrical and (b) channel waveguide. The position and orientations of the dipoles are detailed in the main text.}
\label{totalGvsplasmonicG}
\end{figure}

For high $\beta$ factor, a dipole couples mainly to plasmon modes and this, in turn, warrants that the qubit-qubit interaction is predominantly plasmon-assisted. Under this condition, Eqs.~(\ref{gij}) and (\ref{gammaij}) for $g_{ij}$ and $\gamma_{ij}$ can be evaluated using the plasmonic contribution of the Green's tensor, $\mathbf{G}_{\textrm{pl}}(\mathbf{r},\mathbf{r}')$, of Eq.~(\ref{Gplasmon}) instead of the total one, $\mathbf{G}(\mathbf{r},\mathbf{r}')$. The resulting approximations for the dipole-dipole shift and decay rates are as follows\cite{martincano10}:
\begin{eqnarray}
\label{approximations1}
g_{ij}\simeq g_{ij,\,\textrm{pl}}=\frac{\gamma}{2} \beta e^{-d/2\ell}\sin(k_{\textrm{r}}d)\\
\label{approximations2}
\gamma_{ij}\simeq \gamma_{ij,\,\textrm{pl}}=\gamma \beta e^{-d/2\ell}\cos(k_{\textrm{r}}d),
\end{eqnarray}
where it has been assumed that the transverse position of both qubits and their orientations are identical. Notice that plasmonic decay is accounted for in Eqs.~(\ref{approximations1}) and (\ref{approximations2}) by the presence of the exponential factor $e^{-d/2\ell}$. In order to check the validity of this approximation a comparison of the exact parameters ($g_{ij}$, $\gamma_{ij}$) and the approximate ones ($g_{ij,\,\textrm{pl}}$, $\gamma_{ij,\,\textrm{pl}}$) is presented in Fig.~\ref{totalGvsplasmonicG} for the cylinder [panel (a)] and the channel [panel (b)]. All parameters are normalized to the vacuum decay rate $\gamma_0$. In both cases, the position and orientation of the qubits are chosen to maximize $\beta$, \emph{i.e.}, $h=20\,\textmd{nm}$ and vertical orientation for the cylinder, and $h=150\,\textmd{nm}$ and horizontal orientation for the channel. The parameters are represented as a function of the qubit-qubit separation, $d$, normalized to the modal wavelength, $\lambda_{\textrm{pl}}=2\pi/k_{\textrm{r}}$ (at the operating wavelength $\lambda_{\textrm{pl}}$ is $417\,\textmd{nm}$ for the cylinder and $474\,\textmd{nm}$ for the channel). As expected, the approximation is good for the cylinder and excellent for the channel, in consonance with the corresponding $\beta$ factors (0.6 and 0.9, respectively). For the cylinder, at the chosen $h$, the radiative modes play a small but non-negligible role which shows up as a small disagreement between the exact and approximate results. For both PWs and very small $d$, many radiative and guided modes contribute to the interaction and the approximation breaks down. A different approach to this issue leading to the same result can be found in Ref.~\onlinecite{dzsotjan}. The coupling parameters $g_{ij}$ and $\gamma_{ij}$ are functions of the separation $d$ which oscillate with a periodicity given by the plasmonic wavelength, $\lambda_{\textrm{pl}}$, and decay exponentially due to the ohmic absorption of the plasmonic mode. Notice that the maxima of $\gamma_{ij}$ and those of $g_{ij}$ are shifted a distance $\lambda_{\textrm{pl}}/4$, which implies that the noncoherent and coherent terms of the master equation have different weights for different qubit-qubit separations, a fact that will be important in Sec.~V.

\subsection{Dipoles with different orientations or vertical positions}
\begin{figure}[t]
\begin{center}
\includegraphics[width=0.99\linewidth,angle=0]{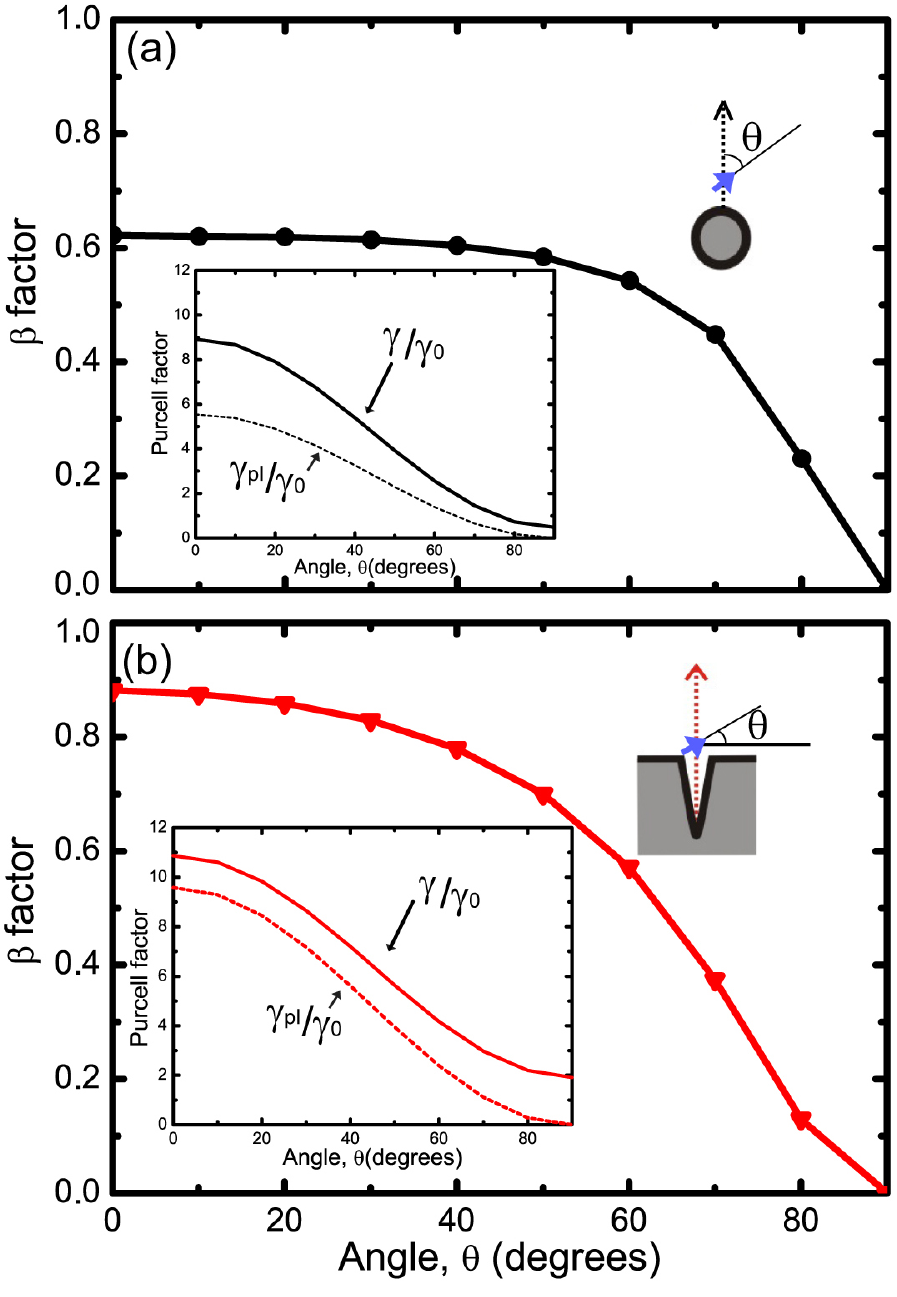}
\end{center}
\caption{ (Color online) Beta factor of one emitter as a function of the angle, $\theta$, formed by the electric field and the dipole moment. (a) Cylinder, and (b) channel. The insets show the total (continuous line) and plasmon (dashed line) decay rates normalized to the vacuum decay rate, ($\gamma/\gamma_0$, $\gamma_{\textrm{pl}}/\gamma_0$), as a function of $\theta$. The positions of the dipoles are detailed in the main text.}
\label{angledependence}
\end{figure}

To close the analysis of the coupling parameters, we now discuss the case when the dipoles have different dipole moment orientations or vertical locations. This is very important from the experimental point of view since a controlled positioning of the emitters is technically challenging\cite{englund10,kramer10,huck11}. When the two dipoles are inequivalent in orientation or position, the mutual decay rates are obtained in a similar way than Eq.~(\ref{approximations2}) and can be expressed as
\begin{equation}
\label{misalignements}
\gamma_{ij,\,\textrm{pl}}=\sqrt{\gamma_{ii}\gamma_{jj}}\sqrt{\beta_{i}\beta_{j}} e^{-d/2\ell}\cos(k_{\textrm{r}}d),
\end{equation}
which indicates that $\beta$´s and $\gamma$´s of both dipoles should be as high as possible to obtain a high $\gamma_{ij,\,\textrm{pl}}$. Since the dependence of $\beta$ with the vertical distance has been discussed already, we focus now on the case of identical transverse positions but different orientations for the dipoles. The dependence of $\beta$ with the angular deviation of the dipole with respect to the electric field polarization is illustrated in Fig.~\ref{angledependence}. Panels (a) and (b) correspond to the cylinder and the channel, respectively. In both cases the emitter position is chosen to maximize $\beta$ ($h=20\,\textmd{nm}$ for the cylinder, and $h=150\,\textmd{nm}$ for the channel). The dipole moment is parallel to the transverse plane, and the definitions of the angular deviation, $\theta$, are sketched in the diagrams of the corresponding panels. As a general rule, the deviation of the dipole from the electric field direction has a detrimental effect, and $\beta$ becomes null for $\theta=90^{\circ}$. Nevertheless, there is a broad angular range where $\beta$ remains relatively stable so that it is not critically affected by relatively large misalignements. Figure~\ref{angledependence} shows that $\beta$ deviates less than a 10\% of the maximum value within a $\theta$-range of $\Delta \theta=60^{\circ}$ for cylinders and of $\Delta \theta=40^{\circ}$ for channels. The functional dependence of $\beta$ with $\theta$ is not simple because although $\gamma_{\textrm{pl}}\propto \cos\theta$ [see Eq.~(\ref{gamma-pl})], $\gamma$ has a more complex dependence. This can be observed by comparison of the curves in the insets of Fig.~\ref{angledependence}. We conclude this section with a brief summary of its main results. We have derived simplified expressions for $g_{ij}$ and $\gamma_{ij}$, which depend on $\beta$ and $\gamma$, and the analysis has shown that channel PWs display higher values of the later parameters. Therefore, to achieve a larger qubit-qubit coupling, we will mainly focus on channel waveguides in the discussion of the generation of entanglement in the next section.

\section{Entanglement generation}
\subsection{Spontaneous decay of a single excitation}
\begin{figure}[b]
\begin{center}
\includegraphics[width=0.99\linewidth,angle=0]{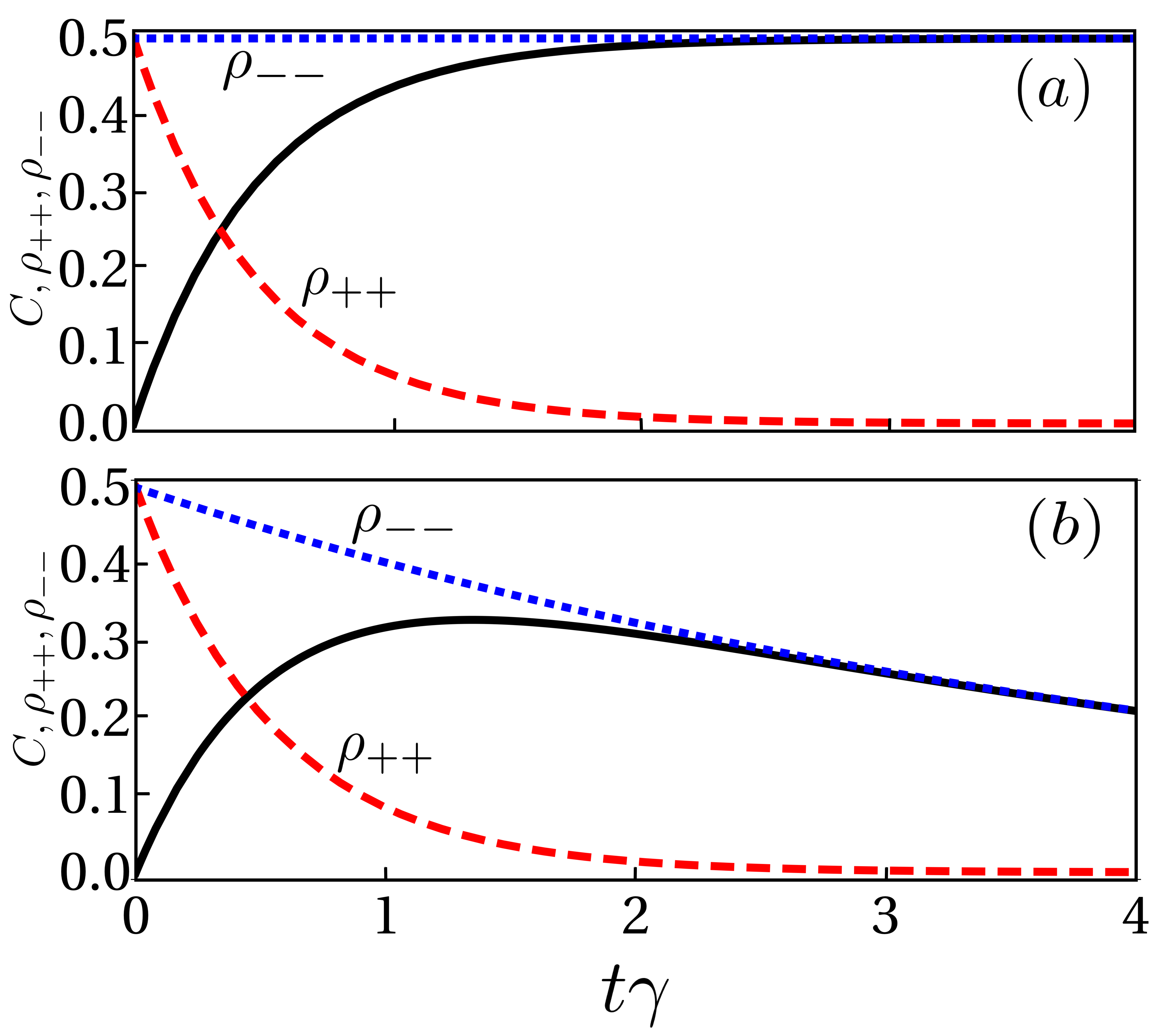}
\end{center}
\caption{ (Color online) Concurrence (black thick line) and populations $\rho_{++}$ (red dashed line), $\rho_{--}$ (blue dotted line), versus time. (a) Ideal PW satisfying $\beta=1$ and $\ell=\infty$. (b) Realistic channel PW. The time is scaled with the emitter lifetime (1/$\gamma$).}
\label{concurrence-nopumping}
\end{figure}

We first consider two identical qubits in front of a channel PW without external pumping. The qubits separation is set as $d=\lambda_{\textrm{pl}}$ and their transverse positions and orientations are identical and chosen to maximize the $\beta$ factor. In this simple but insightful configuration, $g_{ij}$ vanishes and $\gamma_{ij}$ attains its maximum value [Fig.~\ref{totalGvsplasmonicG}(b)], which means that the two-qubit dynamics is purely dissipative. The system is initialized in the (unentangled) state $\ket{1}=\ket{e_1 g_2}=\frac{1}{\sqrt{2}}(\ket{+}+\ket{-})$. In this case the evolution is confined to the subspace spanned by $\{\ket{0}, \ket{+}, \ket{-}\}$ and the master equation is reduced to
\begin{eqnarray}
\label{simplemastereq}
\nonumber\dot{\rho}_{++}(t)&=&-(\gamma+\gamma_{12})\rho_{++}(t)\\
\nonumber\dot{\rho}_{--}(t)&=&-(\gamma-\gamma_{12})\rho_{--}(t)\\
\nonumber\dot{\rho}_{00}(t)&=&(\gamma+\gamma_{12})\rho_{++}(t)+(\gamma-\gamma_{12})\rho_{--}(t),\\
\nonumber\dot{\rho}_{+-}(t)&=&-\gamma\rho_{+-}(t).\\
\end{eqnarray}
There are only a few non-zero entries in $\mathbf{\hat{\rho}}(t)$ and the resulting expression for the concurrence is very simple:
\begin{equation}
\label{simpleconcurrence}
C(t)=\sqrt{[\rho_{++}(t)-\rho_{--}(t)]^2+4\textrm{Im}[\rho_{+-}(t)]^2},
\end{equation}
where we see that an imbalance of the populations $\rho_{++}$ and $\rho_{--}$ results in a non-zero concurrence ($\rho_{+-}(t)$ is real for the chosen conditions). Solving Eq.~(\ref{simplemastereq}), the concurrence becomes
\begin{equation}
\label{supersimpleconcurrence}
C(t)=e^{-\gamma t}\sinh{[\gamma \beta e^{-\lambda_{\textrm{pl}}/(2\ell)}t]}.
\end{equation}
This concurrence and the relevant populations are plotted in Fig.~\ref{concurrence-nopumping} as a function of time ($C$ is the black thick line, and $\rho_{++}$, $\rho_{--}$ are the red dashed and blue dotted lines, respectively). Panel~(a) corresponds to the idealized case where $\beta=1$ and the plasmon propagation length is $\ell=\infty$. The entanglement grows with time monotonically up to a value of $C=0.5$. This process can be easily understood using Eq.~(\ref{simpleconcurrence}) and observing the mentioned population imbalance. Since $\gamma_{12}=\gamma$, the population $\rho_{++}$ decays at an enhanced rate $2\gamma$, whereas $\rho_{--}$ stays constant due to its zero decay rate. Panel~(b) corresponds to a realistic channel PW with $\beta=0.9$ and $\ell=1.7\,\mu\textrm{m}$. In this case the concurrence reaches a maximum value of $C=0.33$ for $t\simeq 1/\gamma$ and then decays exponentially to zero. Again, the entanglement generation is a consequence of the populations imbalance. For this realistic structure both populations have finite decay rates and the concurrence eventually vanishes. The same setup with a cylindrical waveguide produces qualitatively similar results as in Fig.~\ref{concurrence-nopumping}(b) but, since $\beta=0.6$ in this case, the maximum of the concurrence is lower, $C=0.21$. In all three cases, $\ket{+}$ and $\ket{-}$ are examples of superradiant and subradiant states, respectively. We can now present a qualitative picture of more general entanglement generation processes by referring to Fig.~\ref{levelscheme}. The upper level depopulates along two routes: through the state $\ket{+}$, with decay rate $\gamma+\gamma_{12}$, and through the state $\ket{-}$ with decay rate $\gamma-\gamma_{12}$. It is the difference in the decay rates along both routes what results in the transient build up of the concurrence. Notice that the magnitude and sign of $\gamma_{12}$ depend on $d$ (Fig.~\ref{totalGvsplasmonicG}) causing that the roles of the states $\ket{+}$ and $\ket{-}$ are exchanged for $d=\lambda_{\textrm{pl}}/2$, $\ket{+}$ being subradiant and $\ket{-}$ superradiant.

\subsection{Stationary state under external continuous pumping}
\begin{figure}[t]
\begin{center}
\includegraphics[width=0.99\linewidth,angle=0]{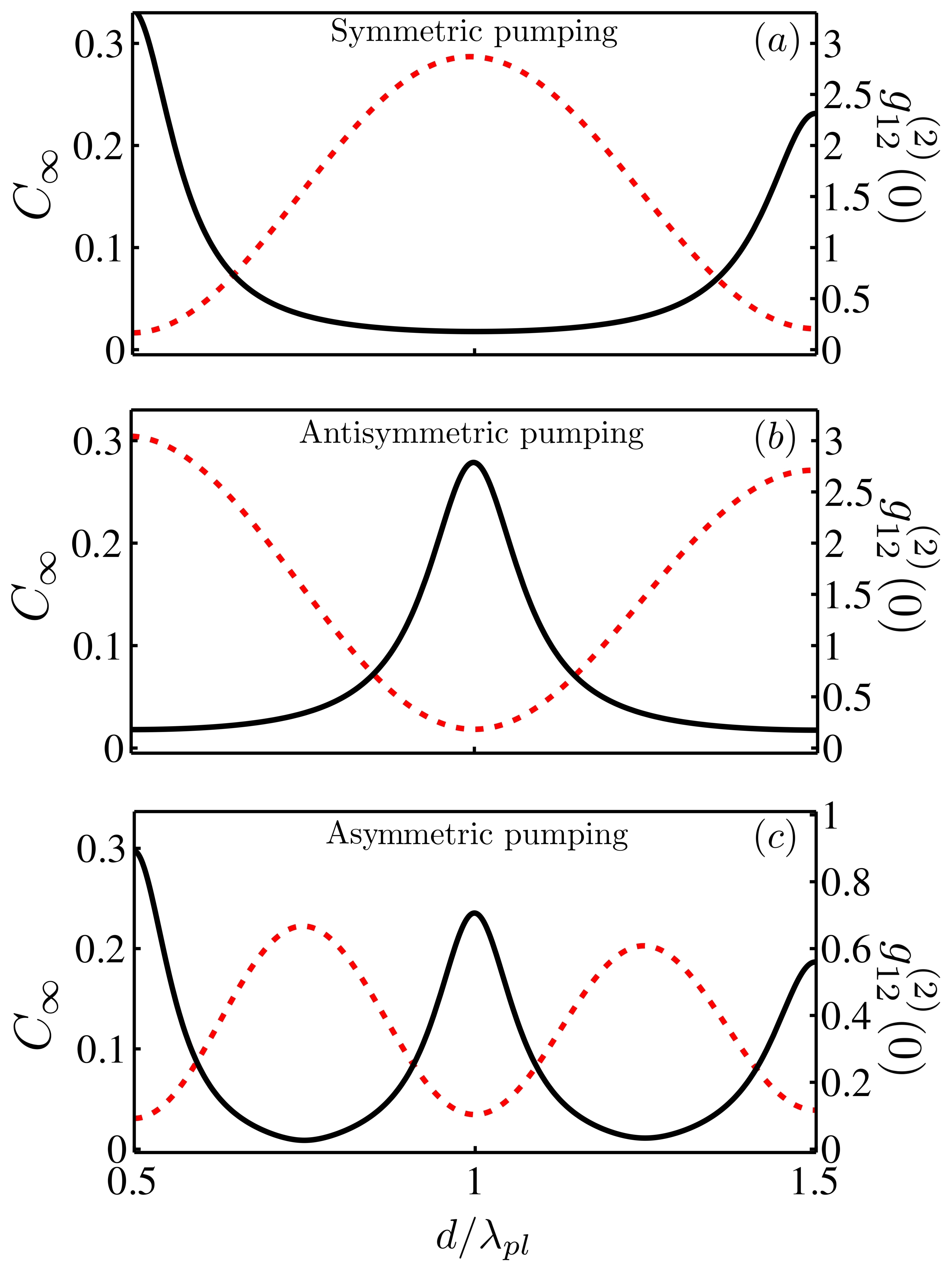}
\end{center}
\caption{(Color online) Steady state concurrence (black line) and qubit-qubit correlation (red dashed line) as a function of the normalized separation $d/\lambda_{\textrm{pl}}$. (a) Symmetric pumping ($\Omega_1=\Omega_2=0.1\gamma$), (b) antisymmetric pumping ($\Omega_1=-\Omega_2=0.1\gamma$), and (c) asymmetric pumping ($\Omega_1=0.15\gamma,\,\Omega_2=0$).}
\label{concurrence-steadystate}
\end{figure}

We have just seen the spontaneous generation of entanglement but, as explained above, the process is a transient phenomenon. To compensate the depopulation of the upper levels, the system could be externally pumped by means of a laser in resonance with the frequency of the qubits\cite{gonzaleztudela11a,gonzaleztudela11b}. The concurrence reached in the corresponding steady state, $C_{\infty}$, is plotted in Fig.~\ref{concurrence-steadystate} (black lines) as a function of the qubits separation normalized to the modal wavelength, $d/\lambda_{\textrm{pl}}$. Three kinds of coherent driving have been considered, differing in the relative phase of the laser fields acting on qubit 1 and 2: symmetric pumping means identical Rabi frequencies, $\Omega_1=\Omega_2$ [panel~(a)], antisymmetric pumping means $\Omega_1=-\Omega_2$ [panel~(b)], and asymmetric pumping corresponds to $\Omega_1\neq0,\,\Omega_2=0$ [panel~(c)]. The absolute value of the non-zero Rabi frequencies is $0.15\gamma$ for the asymmetric pumping and $0.1\gamma$ for the other two situations, \emph{i.e.}, relatively weak. It is very important to realize that we consider now arbitrary separations between the qubits and this implies that both coherent and dissipative dynamics are active, its relative weight depending on $d$ (Fig.~\ref{totalGvsplasmonicG}). With independence of the pumping scheme the concurrences $C_{\infty}$ in Fig.~\ref{concurrence-steadystate} present an oscillating behavior with the qubits separation, and damped due to the plasmon absorption. Importantly, the concurrence maxima occur for those $d/\lambda_{\textrm{pl}}$ where the absolute value of $\gamma_{ij}$ is maximum (Fig.~\ref{totalGvsplasmonicG}). This suggests a relationship between entanglement generation and dissipative two-qubit dynamics. Let us justify the position of the maxima of $C_{\infty}$ applying the ideas developed for the undriven case. When the pumping is symmetric [panel~(a)], the laser populates the symmetric state $\ket{+}$. This state is subradiant for $d=\frac{1}{2}\lambda_{\textrm{pl}}, \frac{3}{2}\lambda_{\textrm{pl}},\ldots$ leading to a population imbalance and the corresponding concurrence. For $d=\lambda_{\textrm{pl}}, 2\lambda_{\textrm{pl}},\ldots$, $\ket{+}$ is superradiant and the pumping is not able to induce a significant $\rho_{++}$ population. For antisymmetric pumping [panel~(b)] it is the state $\ket{-}$ which is populated. This state is subradiant for $d=\lambda_{\textrm{pl}}, 2\lambda_{\textrm{pl}},\ldots$ again leading to a population imbalance and entanglement. For $d=\frac{1}{2}\lambda_{\textrm{pl}}, \frac{3}{2}\lambda_{\textrm{pl}},\ldots$, the situation is reversed. Finally, for asymmetric pumping [panel~(c)] both $\ket{+}$ and $\ket{-}$ are populated and the situation is a mixture of the previous two. In this case maxima are found for $d=\frac{1}{2}\lambda_{\textrm{pl}},\lambda_{\textrm{pl}},\frac{3}{2}\lambda_{\textrm{pl}},\ldots$, their concurrence being slightly smaller than that found for the symmetric or antisymmetric pumping.

\begin{figure}[htbp]
\begin{center}
\includegraphics[width=0.99\linewidth,angle=0]{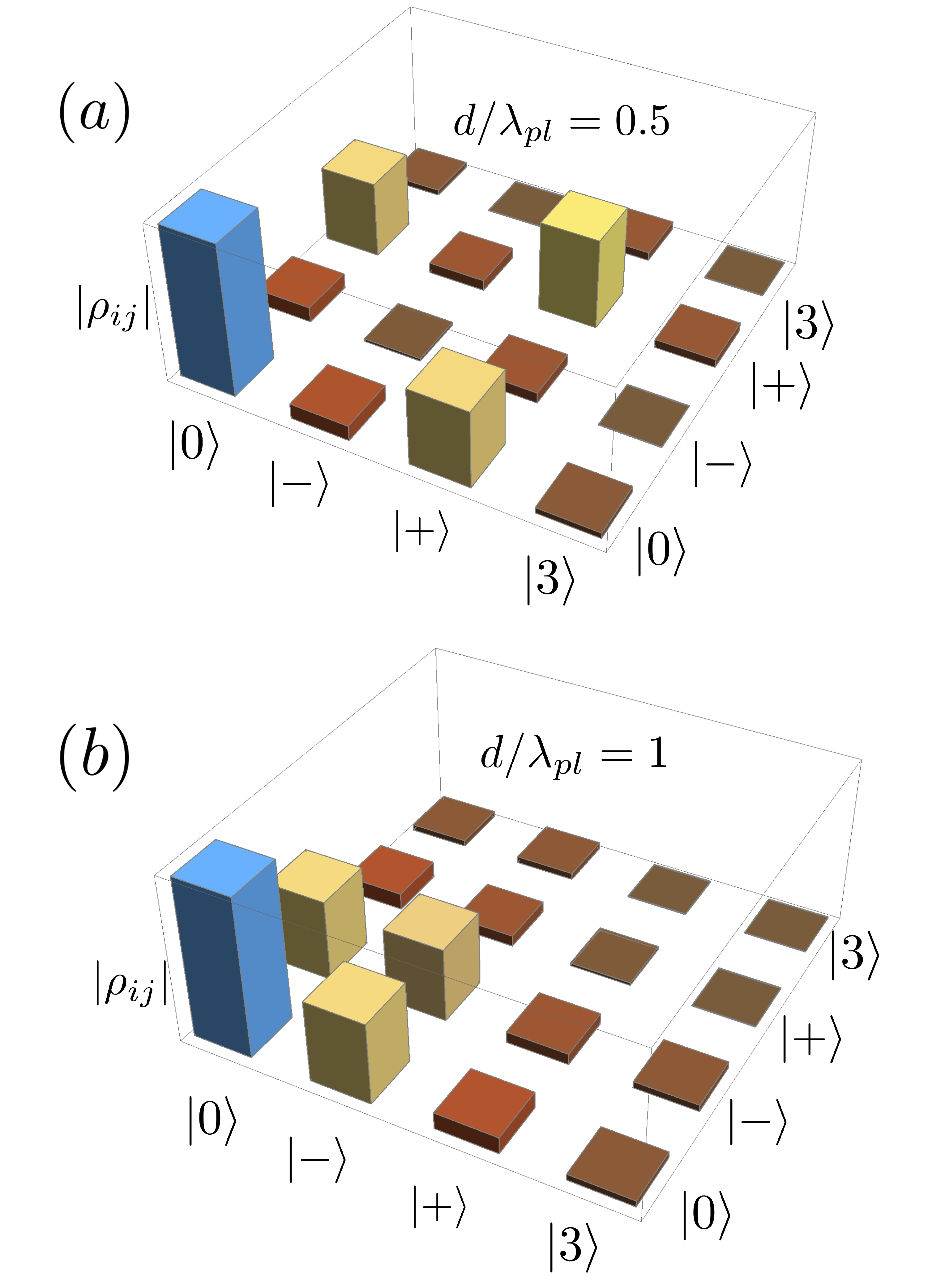}
\end{center}
\caption{Tomography of the absolute value of the elements of the steady state density matrix for asymmetric pumping ($\Omega_1=0.15\gamma,\,\Omega_2=0$). (a) $d=\lambda_{\textrm{pl}}/2$, and (b) $d=\lambda_{\textrm{pl}}$.}
\label{tomography}
\end{figure}

To verify that the previous interpretation is correct, we plot the tomography of the steady state density matrix in Fig.~\ref{tomography}. We choose the case of asymmetric pumping and two different qubit separations. In panel (a) $d=\lambda_{\textrm{pl}}/2$ and, besides the population of the ground state, we recognize the large $\rho_{++}$ population of the subradiant state $\ket{+}$ driven by the pumping, and the negligible $\rho_{--}$ population of the superradiant state $\ket{-}$. For $d=\lambda_{\textrm{pl}}$ [panel~(b)], we now observe a large $\rho_{--}$ population of the subradiant state $\ket{-}$ driven by the pumping, and a negligible $\rho_{++}$ population of the superradiant state $\ket{+}$. Let us remark that, strictly speaking,  Eq.~(\ref{simpleconcurrence}) is not correct when pumping is included, because now further elements of $\mathbf{\hat{\rho}}$ are non zero. However, the tomography shows that these additional elements are very small and Eq.~(\ref{simpleconcurrence}) should be approximately valid, justifying the argument that population imbalance leads to concurrence. Since this population imbalance is due to the different decay rates of the super- and subradiant states, both of which are produced by dissipation, we want to emphasize that the entanglement generation is driven by the two-qubit dissipative dynamics. At this point, a brief comparison with the results that can be achieved with cavities may be useful again. In cavity QED there is mainly coherent coupling between the qubits but no cross decay, and a coherently pumped cavity is unable to generate any significant concurrence. It would be possible to work with an incoherent pumping with cross terms\cite{delValle07,delValle11a} but, as mentioned previously, this scheme is experimentally more difficult than our proposal.

Once the tomography of the density matrix is known, the calculation of concurrence (or any other equivalent entanglement quantifier) is straightforward. However, tomographic procedures are experimentally cumbersome and, for this reason, it is of interest to establish connections between entanglement and other more easily measurable magnitudes. In our two-qubit system, entanglement is associated with the probability that the state of the system is $\ket{+}$ or $\ket{-}$. In other words, entanglement is related with having a strong correlation between the states $\ket{1}=\ket{e_1 g_2}$ and $\ket{2}=\ket{g_1 e_2}$. This must manifest in the correlation between one photon emitted from qubit 1 and another photon emitted from qubit 2. Hanbury Brown-Twiss-like experiments are able of measuring photon-photon correlations and, in particular, the cross-term of the second order coherence function which, for zero delay, takes the form \cite{walls94,scully_book02}
\begin{eqnarray}
\label{g2}
g^{(2)}_{12}=\frac{<\sigma ^\dagger _1 \sigma ^\dagger _2 \sigma _2 \sigma  _1 >}{<\sigma ^\dagger _1 \sigma  _1 ><\sigma ^\dagger _2 \sigma _2 >}.
\end{eqnarray}
Figure~\ref{concurrence-steadystate} displays together the concurrence $C_{\infty}$ (black continuous lines) and the second order correlation function at zero delay $g^{(2)}_{12}$ (red dashed lines). In all three panels it is observed that when $C_{\infty}$ is large, a clear antibunching ($g^{(2)}_{12} \rightarrow 0$) takes place, which is consistent with the system predominantly being in a state $\ket{+}$ or $\ket{-}$. On the other hand, when $C_{\infty} \rightarrow 0$, $g^{(2)}_{12}$ grows and the antibunching is reduced, which is again consistent with a decreased correlation between $\ket{1}$ and $\ket{2}$. The main result to be drawn is the distinct relationship between $C_{\infty}$ and $g^{(2)}_{12}$. Lacking an analytical expression relating $C_{\infty}$ and $g^{(2)}_{12}$, our results clearly support the idea of measuring cross-terms of the second order coherence, at zero delay, as a manifestation of entanglement.

\begin{figure}[t]
\begin{center}
\includegraphics[width=0.99\linewidth,angle=0]{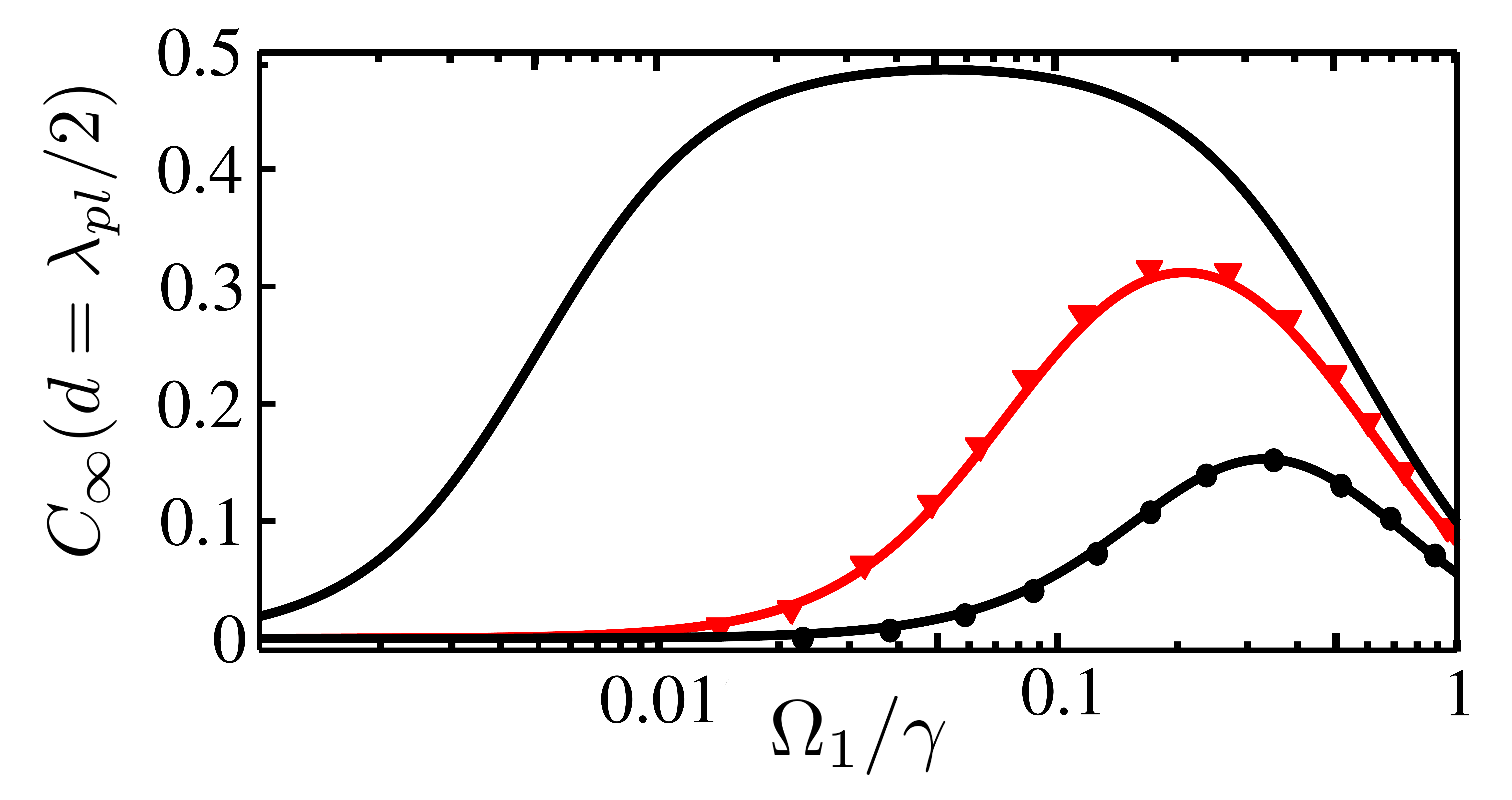}
\end{center}
\caption{ (Color online) Steady state concurrence as a function of the driving laser power for asymmetric pumping ($\Omega_1\neq0,\,\Omega_2=0$) and qubits separation $d=\lambda_{\textrm{pl}}/2$. Ideal case $\beta=1$ (black line), cylinder $\beta=0.6$ (black circles), and channel $\beta=0.9$ (red triangles).}
\label{laserpower}
\end{figure}

Up to now we have considered a weak pumping rate. For an experimental implementation of our proposal, it is important to determine the pumping rate range for which the described phenomena may happen. The influence of the pumping intensity is analyzed in Fig.~\ref{laserpower}, which renders $C_{\infty}$ versus $\Omega_1/\gamma$. Here asymmetric pumping is considered and a qubit separation $d=\lambda_{\textrm{pl}}/2$. The results are computed for three waveguides: a cylinder ($\beta=0.6$, black circles), a channel ($\beta=0.9$, red triangles), and an ideal waveguide ($\beta=1$ and no absorption, black line). Each structure presents an optimum pumping power to achieve maximum concurrence. In order to obtain a non-negligible concurrence, the subradiant state has to be populated at a rate faster than its lifetime, which explains both why concurrence is small at low pumping rates and why the structures with lower $\beta$ require a higher pumping to reach their optimum entanglement.  In addition, we observe that the maximum attainable concurrence improves for higher $\beta$ factor, which again justifies the use of channel instead of cylindrical PWs.

\begin{figure}[htbp]
\begin{center}
\includegraphics[width=0.99\linewidth,angle=0]{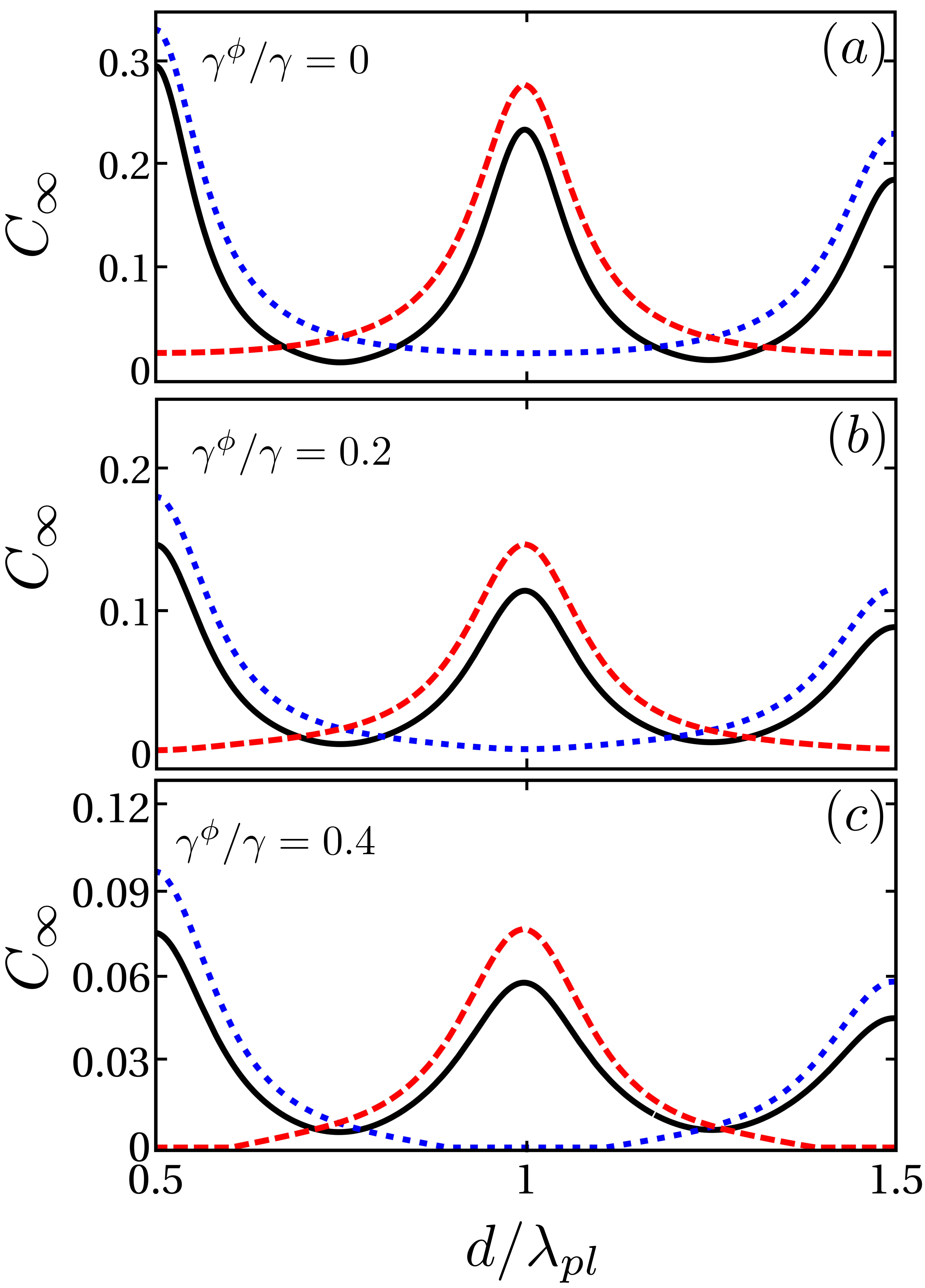}
\end{center}
\caption{(Color online) Steady state concurrence as a function of the normalized separation $d/\lambda_{\textrm{pl}}$ for different pumping conditions and dephasing rates. (a)~$\gamma^\phi/\gamma =0.0$, (b)~$\gamma^\phi/\gamma =0.2$, and (c)~$\gamma^\phi/\gamma =0.4$. In all panels the blue dotted lines correspond to symmetric pumping ($\Omega_1=\Omega_2=0.1\gamma$), the red dashed lines correspond to antisymmetric pumping ($\Omega_1=-\Omega_2=0.1\gamma$), and the black continuous lines correspond to asymmetric pumping ($\Omega_1=0.15\gamma,\,\Omega_2=0$). Notice that the vertical scale is not the same in the three panels.}
\label{dephasing}
\end{figure}

Finally, we pay attention to how the generation of entanglement is affected by the presence of dephasing. For this purpose we have recomputed the dynamics of the system including now in the master equation (\ref{masterequation}) an additional term representing pure dephasing. This term is given by\cite{delValle11a}
\begin{equation}
\label{dephAgarwal}
  {\mathcal L}_\mathrm{deph}[\hat{\rho}]=\frac{\gamma^\phi}{2}\sum_{i}\Big[[\hat{\sigma}_i^\dag \hat{\sigma}_i,\hat{\rho}],\hat{\sigma}_i^\dag \hat{\sigma}_i\Big].
\end{equation}
The value of the dephasing rate $\gamma^\phi$ is difficult to estimate in general because it is very dependent on the particular realization of the qubit and it is strongly influenced by the presence of the metallic part of the system. For nitrogen-vacancy centers in diamond under resonant pumping, pure dephasing times up to $100\,\textmd{ns}$ have been measured\cite{jelezko06,batalov08}. For the typically considered situation in our system, where the Purcell factor is about 10, this corresponds to $\gamma^\phi$ about one hundredth of the emission rate $\gamma$. In our calculations we will consider larger dephasing values, both as a conservative measure and because they may be more relevant for other emitter types. Figure~\ref{dephasing} shows the steady state concurrence as a function of the qubit-qubit separation $d$ for different values of the pure dephasing rate and various pumping conditions. Dephasing grows from zero in panel (a) to $\gamma^\phi/\gamma =0.4$ in panel (c). The qualitative behavior is the same in all panels but the value of $C_{\infty}$ decreases as the dephasing rate grows (notice that the vertical scale is not the same in all panels). Nevertheless, the value of the concurrence maxima are non-negligible even in the worst case of panel (c). Moreover, this decrease can be partially compensated by increasing the intensity of the pumping laser. Therefore, our results show that pure dephasing reduces qubit-qubit entanglement but not as much as to preclude its formation by the mediation of the surface modes supported by 1D plasmonic waveguides.

\section{Conclusions}
We have presented a detailed analysis of how plasmonic waveguides can be used to achieve a high degree of entanglement between two distant qubits. A full account of the theoretical framework has been also described. Importantly, the degrees of freedom associated with the surface plasmons can be traced out, leading to a master equation formalism for the two qubits in which the two contributions to the effective interaction between them (coherent and dissipative terms) are then obtained by means of the \emph{classical} electromagnetic Green's function. We have shown that the main ingredients to obtain a high value for the concurrence are a large $\beta$-factor and the one-dimensional character of the surface modes supported by the plasmonic waveguide. By studying how steady-state entanglement can be generated, we have demonstrated that the dissipative part of the qubit-qubit interaction mediated by plasmons is the main driving force in order to achieve entanglement. We have also analyzed the sensitivity of this plasmon-mediated entanglement to different parameters, such as the dipole orientations of the qubits, the pumping rate, and the inherent presence of dephasing mechanisms in the system. In all cases, we have found that the dissipation-driven generation of entanglement is robust enough to be observed experimentally by using plasmonic waveguides that are currently available. Finally, we have proposed a feasible way to measure the emergence of entanglement in these structures by establishing a direct link between the concurrence and the cross-term second order coherence functions that can be extracted from the experiments. We would like to emphasize that the scheme presented in this paper could be also operative for other types of photonic waveguides provided that the two main ingredients described above (large $\beta$-factor and quasi-1D character) are present. Our results demonstrate that plasmonic waveguides can be used as a reliable tool-box for studying and devising quantum optics phenomena without the necessity of a cavity.

\acknowledgments
Work supported by the Spanish MICINN (MAT2008-01555, MAT2008-06609-C02, CSD2006-00019-QOIT and CSD2007-046-NanoLight.es) and CAM (S-2009/ESP-1503).
D.M.-C. and A.G.-T acknowledge FPU grants (AP2007-00891 and AP2008-00101, respectively) from the Spanish Ministry of Education.

\end{document}